\documentclass[aps,prl,reprint,groupedaddress]{revtex4-1}

\usepackage{hyperref}
\usepackage{epsfig}
\usepackage{graphicx}
\usepackage{subfigure}
\usepackage{latexsym}
\usepackage{color}
\usepackage{fullpage}
\usepackage{epstopdf}
\usepackage{dcolumn}
\usepackage{bm}
\usepackage{ulem}
\usepackage{units}

\newcommand{\new}[1]{\textcolor{black}{#1}}
\newcommand{\proofs}[1]{\textcolor{black}{#1}}

\begin{document}

\title{Non-universality of artificial frustrated spin systems}

\author{I. A. Chioar$^{1,2}$, N. Rougemaille$^{1,2}$, A. Grimm$^{1,2}$, O. Fruchart$^{1,2}$, E. Wagner$^{1,2}$, M. Hehn$^3$, D. Lacour$^3$, F. Montaigne$^3$, B. Canals$^{1,2}$}

\address{$^1$CNRS, Inst NEEL, F-38042 Grenoble, France\\$^2$ Univ. Grenoble Alpes, Inst NEEL, F-38042 Grenoble, France\\$^3$Institut Jean Lamour, Universit\'e de Lorraine \& CNRS, Vandoeuvre l\`es Nancy, F-54506, France}

\date{\today}

\begin{abstract}
Magnetic frustration effects in artificial kagome arrays of nanomagnets with out-of-plane magnetization are investigated using Magnetic Force Microscopy and Monte Carlo simulations. Experimental and theoretical results are compared to those found for the artificial kagome spin ice, in which the nanomagnets have in-plane magnetization. In contrast with what has been recently reported, we demonstrate that long range (i.e. beyond nearest-neighbors) dipolar interactions between the nanomagnets cannot be neglected when describing the magnetic configurations observed after demagnetizing the arrays using a field protocol. As a consequence, there are clear limits to any universality in the behavior of these two artificial frustrated spin systems. We provide arguments to explain why these two systems show striking similarities at first sight in the development of pairwise spin correlations.
\end{abstract}

\pacs{75.10.Hk, 75.50.Lk, 75.70.Cn, 75.60.Jk}

\maketitle

Frustration is a ubiquitous concept in physics. In some cases, frustration can lead to an extensively degenerate ground state of the considered system. Pauling's description of the low-temperature proton disorder in water ice is probably the first example of frustration in condensed matter physics, and remains its paradigm \cite{Pauling1935}. At the end of the nineties, new magnetic compounds have been synthesized in which the disorder of the magnetic moments at low temperatures is analogous to the proton disorder in water ice \cite{Harris1997}. Since then, intense work has been devoted to these frustrated spin systems \cite{Gingras2009}. This correspondence between the physics of water ice and its magnetic counterparts has been recently extended to artificial realizations of frustrated spin systems \cite{Davidovic1996, Hilgenkamp2003, Wang2006, Libal2006, Han2008}. 

\begin{figure}
\includegraphics[width=6.5cm]{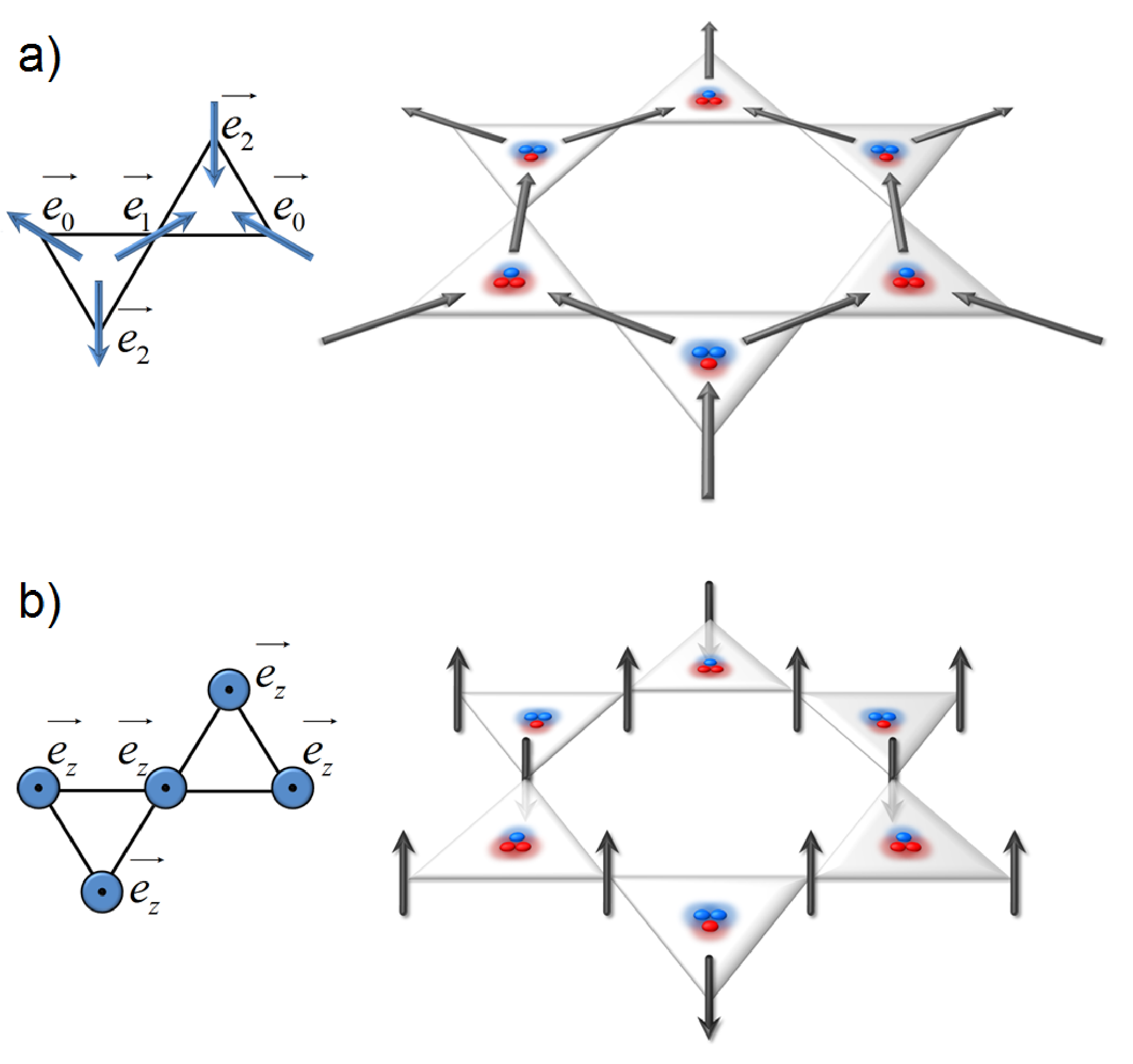}
\caption{\label{fig1}
Sketches of (a) the multiaxial, ferromagnetic kagome spin ice (ksi) model and of (b) the uniaxial, antiferromagnetic kagome Ising (kI) model. The unit vectors $\vec{e}_i$ defining the directions of each spin are represented as blue arrows in the left corner of the two sketches. As a convention for the ksi model, we consider unit vectors pointing outwards (inwards) from a $\bigtriangledown$-type ($\bigtriangleup$-type) triangle. The +/- magnetic charges associated with the dumbbell description of the spins are represented as red/blue clouds, respectively.}
\end{figure}

For \proofs{lithographically-patterned}, two dimensional arrays of nanomagnets, magnetic imaging techniques were successfully used to observe, in real space, how each individual spin of the array locally accommodates frustration \cite{Wang2006, Tanaka2006, Qi2008}, how the entire lattice approaches the ground state manifold \proofs{\cite{Morgan2011, Budrikis2011, Rougemaille2011, Budrikis2012a, Budrikis2012b, Farhan2013a, Farhan2013b, Zhang2013, Montaigne2014}} and how monopole-like excitations form \cite{Ladak2010, Mengotti2011, Phatak2011}. Besides magnetic imaging, artificial spin systems offer the opportunity to change the geometry of the array at will and to explore new phenomena by tuning the (micro)magnetic properties of the nanomagnets \cite{Ladak2011, Zeissler2013, Rougemaille2013}. 

\begin{figure*}
\includegraphics[width=16.5cm]{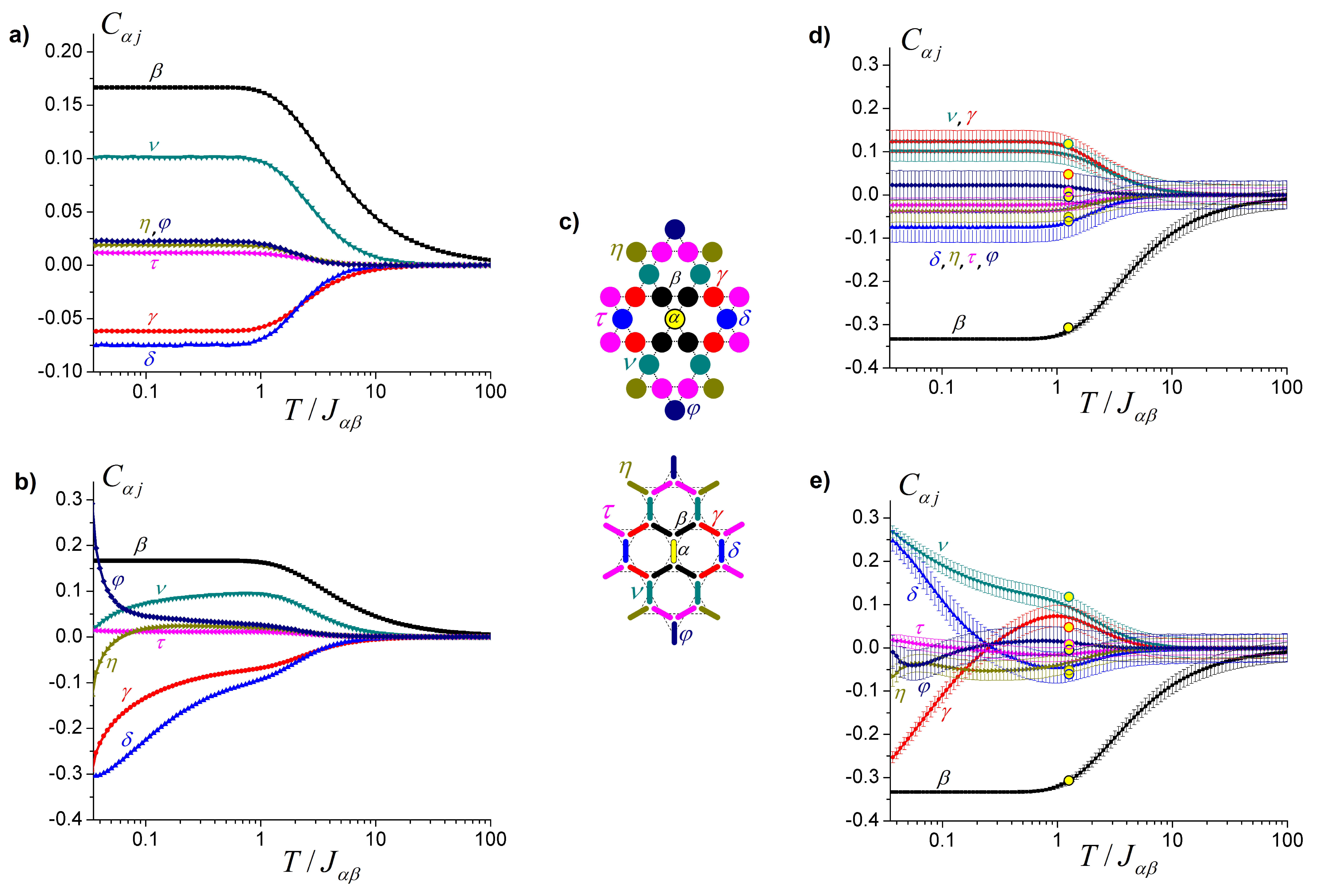}
\caption{\label{fig2}
(Color online) (a-b) Temperature dependence of the theoretical spin correlations for the short range (a) and the long range (b) kagome spin ice models. (c) Definitions and relative indexes for the first seven nearest neighbors that we consider in this work. (d-e) Temperature dependence of the theoretical spin correlations for the short range (d) and the long range (e) kagome Ising models. The yellow circles are the experimental values of the corresponding spin correlators \proofs{extracted from the final magnetic configuration corresponding to one of our MFM images (image index 5 - see Figure 6)}.}
\end{figure*}

So far, most efforts have been focused on square and kagome lattices of in-plane magnetized nanomagnets. However, Zhang and coworkers have recently investigated the properties of an artificial frustrated spin system, in which the nanomagnets have out-of-plane magnetization \cite{Zhang2012}. Contrary to other studies in which nanomagnets are coupled both ferromagnetically and antiferromagnetically, depending on the considered pair of spins, in this geometry uniaxial Ising pseudo-spins are all coupled antiferromagnetically through the magnetostatic interaction (see Figure 1). A new artificial spin model was therefore fabricated and its properties were investigated using Magnetic Force Microscopy (MFM) after demagnetizing the system using an AC field protocol. One important conclusion has been drawn from this study that carefully compares the pairwise spin correlations of the multiaxial, ferromagnetic kagome spin ice (ksi) model with those of the uniaxial, antiferromagnetic kagome Ising (kI) model: the two systems (see Fig.1), described by spin models based solely on nearest-neighbor interactions, show striking similarities in the development of moment pair correlations, indicating a universality in artificial spin ice behavior. The physics of field-demagnetized artificial spin ice systems thus seems to transcend the particular material realization, and even the geometry of the magnetic moments \cite{Zhang2012}.

Investigating the properties of similar artificial kagome arrays of nanomagnets with out-of-plane magnetization, we end up with a different conclusion: our experimental findings can only be described by spin models that include long range dipolar interactions, breaking the apparent universality between the ksi and kI frustrated systems, as they develop clearly distinctive pairwise spin and charge correlations. These results are of considerable importance since the dipolar interaction lifts the degeneracy of the spin ice manifold and induces new magnetic phases that do not exist in the corresponding short range models \cite{Moller09}.

Comparing the ksi and kI models requires a set of common definitions and conventions.
Interactions between the spins are described by a Heisenberg-like Hamiltonian: $H=-\sum_{i < j}^{} J_{ij} \vec{S}_i. \vec{S}_j$ where $J_{ij}$ is the coupling constant between spins $\vec{S}_i$ and $\vec{S}_j$. The spin vectors can be written as: $\vec{S}_i = \sigma_{i} \vec{e}_i$ where $\vec{e}_i$ is the unit vector that defines the direction of the spin $i$, while $\sigma_{i}$ is a scalar that defines the spin's orientation along this direction ($+1$ if parallel to the unit vector and $-1$ if antiparallel). The fact that $\sigma_{i} = \pm 1$ indicates that the spins are of Ising type. For the ksi model, spins point along the bisectors of the triangles, hence three spin directions are considered. The orientation of the corresponding unit vectors is a matter of convention and we consider the unit vectors pointing outwards from a $\bigtriangledown$-type triangle and therefore inwards in the case of a $\bigtriangleup$-type triangle, as illustrated in Figure 1a. For the kI model, there is only one unit vector $\vec{e}_z$ pointing perpendicular to the kagome plane (Fig. 1b).

With this set of definitions, the Hamiltonians associated to the short range versions (nearest-neighbor interactions only) of these two models can be written respectively as: $H_{ksi}=J / 2 \sum_{i < j}^{} \sigma_i. \sigma_j$ and $H_{kI}=-J^{'} \sum_{i < j}^{} \sigma_i. \sigma_j$. If $J / 2 = -J^{'}$ the two models are identical. In other words, the two models map one-another and must develop identical pairwise spin correlations. Experimentally, if long range interactions can be neglected, it is then expected to observe similarities in the development of moment pairwise correlations when comparing artificial realizations of these two models. However, this mapping is not valid anymore if long range dipolar interactions are taken into account: as previously mentioned, while in the kI model these interactions are isotropic and always favor an antiferromagnetic alignment of the spins for all distances, in the ksi model they lead to an effective ferromagnetic or antiferromagnetic coupling depending on the considered pair of spins.

We performed Monte Carlo simulations for both cases, starting from a high-temperature paramagnetic regime and then sequentially reducing the temperature down to lower energy manifolds. The simulations were done \proofs{on} a network of $18 \times 18 \times 3$ lattice sites (i.e. the typical size of our experimental arrays, see below) with periodic boundary conditions. We used a single spin flip algorithm and a simulated annealing procedure from $T/J_{\alpha \beta} = 100$ to $T/J_{\alpha \beta} = 0.04$, where $J_{\alpha \beta}$ is the coupling constant between nearest-neighbors. In these simulations, 10$^4$ modified Monte Carlo steps are used for thermalization, followed by 10$^4$ modified Monte Carlo steps for sampling. The temperature dependence of the correlation coefficients $C_{ij} = \langle \vec{S}_i.\vec{S}_j \rangle$ between spins $\vec{S}_i $ and $\vec{S}_j$ are reported in Figure 2 up to the 7th neighbor for both the short range (Fig.2d) and long range (Fig.2e) kI models. As expected, there is no difference between the short range ksi (Fig.2a) and kI (Fig.2d) models, given that the -1/2 geometrical factor is implemented for the respective spin correlations ($C_{\alpha \beta}$, $C_{\alpha \gamma}$, $C_{\alpha \tau}$ and $C_{\alpha \eta}$). On the contrary, when long range dipolar interactions are taken into account \cite{Note_Nico}, the ksi and kI models are different (Fig.2b and Fig.2e respectively), and several spin-spin correlation coefficients exhibit clearly distinctive features. For example, as the temperature drops, the second ($C_{\alpha \gamma}$) and fourth ($C_{\alpha \delta}$) neighbor coefficients (see Fig.2c for the definition of pairwise correlations) continuously decrease in the ksi model, while they have non-monotonous variations in the kI model.

\begin{figure}
\includegraphics[width=8cm]{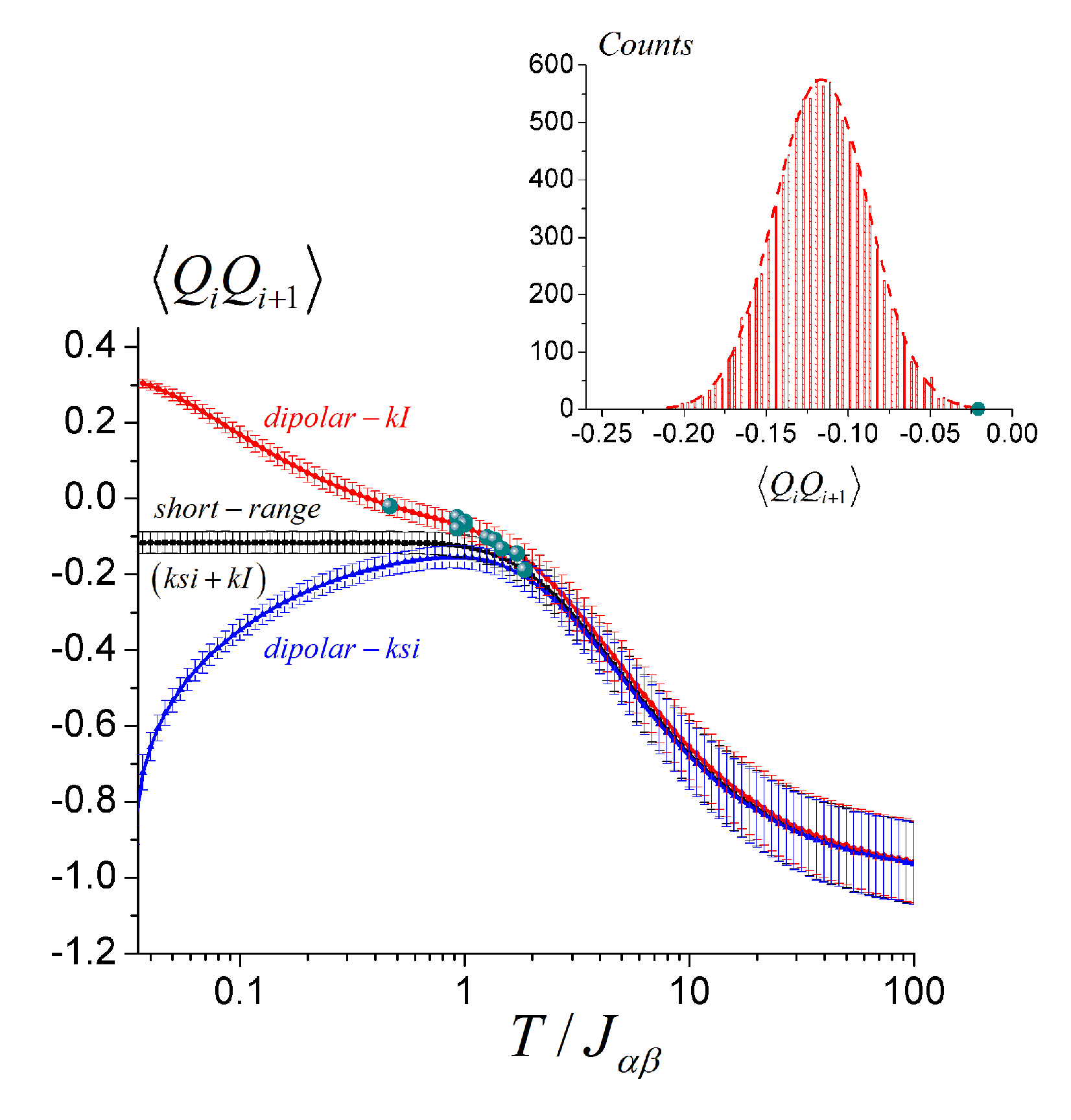}
\caption{\label{fig3}
(Color online) Temperature dependence of the theoretical nearest-neighbor charge correlators and their standard deviations calculated for the short/long range ksi models (black/blue) and the short/long range kI models (black/red). The green circles correspond to the values measured experimentally. Inset shows the histogram of the charge correlator well within the spin ice manifold of the short range kI model together with the value deduced from our measurements.}
\end{figure}

Within the long range interaction picture, the difference between the two models is even more striking when considering the nearest-neighbor charge correlator. In the dumbbell approximation, a spin is treated as a magnetic dipole having two opposite classical magnetic charges. Given our convention (see Fig.1), the charge of vertex $i$ can be written as: $Q_{i}^{\bigtriangleup}= \sum_{k=1}^{3} \sigma_{k}^{i}$ for $\bigtriangleup$-like triangles and $Q_{i}^{\bigtriangledown}= - \sum_{k=1}^{3} \sigma_{k}^{i}$ for $\bigtriangledown$-like triangles. The charge neutrality over the entire lattice is always preserved, regardless of the spin configuration, as each spin contributes with two opposite magnetic charges (see blue and red clouds in Fig.1a). The same definition applies for the kI model. As each spin is the connection point between a $\bigtriangleup$ and a $\bigtriangledown$ triangle, we again define the magnetic charges by summing up the individual spin contributions for every triangle, and we take the $\sigma$ value of the spin in a $\bigtriangleup$ triangle and the $- \sigma$ value in the adjacent $\bigtriangledown$ triangle. All individual contributions are \proofs{thus} taken into account and the charge neutrality condition is intrinsically satisfied (see blue and red clouds in Fig.1b).

The temperature dependence of the charge correlator $\langle Q_i.Q_{i+1} \rangle$ is reported in Figure 3 for the short range and long range ksi and kI models. Since the two short range models are identical, their corresponding charge correlators are the same (black curve). On the contrary, the long range versions of these models exhibit clearly distinctive temperature \proofs{dependencies} for the charge correlator, both in value and sign (see blue and red curves in Figure 3). Interestingly, while it is always negative in the ksi model, the charge correlator becomes positive in the kI model after the system has reached the spin ice manifold. These results \proofs{unambiguously} demonstrate that the two dipolar models are different and develop distinctive moment pair correlations, ruling out the universality concept when interactions beyond nearest-neighbors are taken into account.

The challenge is then to determine whether the physics of artificial realizations of the kI model is governed by short range or long range interactions. To answer this question, kagome arrays of nanodisks have been fabricated from Si//Ta(5nm)/TbCo(40nm)/Ru(2nm) thin films with perpendicular magnetic anisotropy \cite{Gottwald2012}. They have been grown by UHV sputtering, with a base pressure of 10$^{-9}$mbar, by co-sputtering of Co and Tb in DC mode. The power has been adjusted in order to achieve a Tb$_{12}$Co$_{88}$ concentration. The film has been patterned by e-beam lithography and ion beam etching to obtain nanodisks that have a typical diameter of 300 nm and a center to center distance of 400 nm, which ensures that they are physically disconnected and only coupled through the magnetostatic interaction. Due to the system geometry and the perpendicular magnetic anisotropy, the magnetostatic interaction between nearest-neighbors favors an antiferromagnetic alignment of the magnetic elements. We thus made an artificial realization of the kI model. After demagnetizing the arrays using a damped, alternating, out-of-plane magnetic field \cite{Wang2007, Ke2008}, the final magnetic configuration of each nanodisk is resolved by Magnetic Force Microscopy. Typical topographic and magnetic images of the array are shown in Figure 4. Residual magnetization after demagnetization is low, of the order of a few percents, and the kagome ice rule is globally well obeyed ( in general, 3\%-5\% of all the vertices have a 3-in or 3-out spin configuration ).

\begin{figure}
\includegraphics[width=8cm]{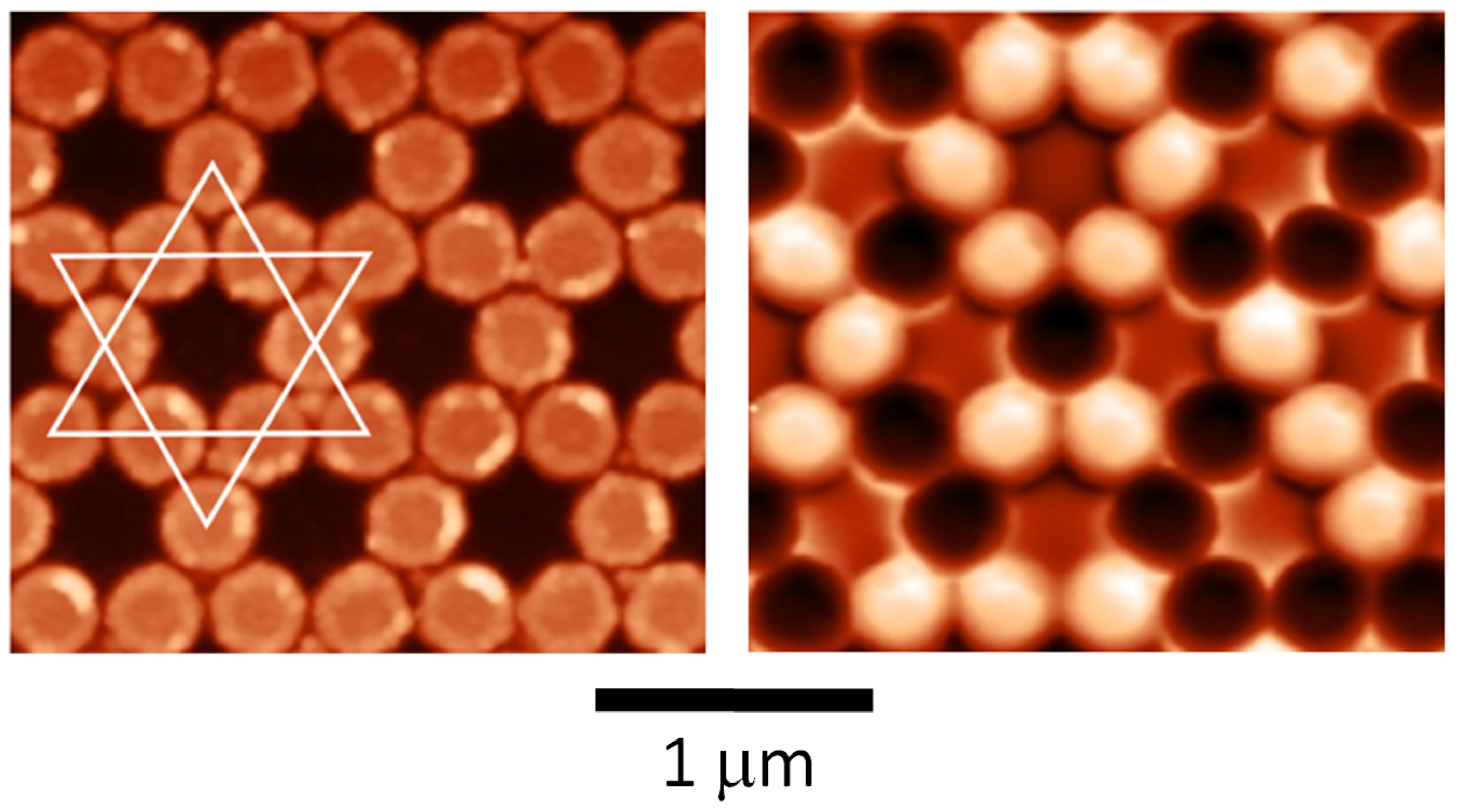}
\caption{\label{fig4}
(Color online) Typical topographic (left) and magnetic (right) images of kagome arrays of TbCo nanodisks with out-of-plane magnetization. The white lines in the topographic image highlights the kagome lattice. Black and white contrast in the magnetic image gives the local direction of magnetization.}
\end{figure}

\proofs{Experimental values for both spin and charge correlators were extracted from 10 different MFM images, each containing about 1000 magnetic elements. Averages performed over one such image for each correlation type determine a set of experimental correlations. A typical set of experimental spin-spin correlations is reported in Figure 2.} To quantify the scattering of our experimental correlations with respect to their corresponding average values given by Monte Carlo simulations, we employed a standard deviation analysis through the use of a "spread-out" function defined as:  $K(T/J_{\alpha \beta})=\sqrt(\sum_{j}^{} (C_{\alpha j}^{exp}-C_{\alpha j}^{MC}(T/J_{\alpha \beta}))^2)$ where $C_{\alpha j}^{exp}$ represent the experimental correlations, while $C_{\alpha j}^{MC}(T/J_{\alpha \beta})$ are the average Monte Carlo correlations at a given temperature $T/J_{\alpha \beta}$, with $j$ ranging from 1, the nearest-neighbor correlation ($C_{\alpha \beta}$), up to 7 ($C_{\alpha \phi}$), and including the nearest-neighbors charge-charge correlations as well \cite{Note_Ioan}. For each set of experimental values, this function can be computed over the entire range of Monte Carlo temperatures for both the short-range and the long-range models (Figure 5).

\new{The minimum of $K(T/J_{\alpha \beta})$ defines an effective temperature for which the optimal fit is achieved. For the experimental values reported in Figure 2, both short and long range models render the same effective temperature, $T/J_{\alpha \beta} = 1.26$. However, the deviations of several spin correlations $(C_{\alpha \beta}, C_{\alpha \gamma}, C_{\alpha \nu})$ are rather high in the short-range picture, exceeding their theoretical standard deviations, whereas the long range model offers a better fit. This aspect is also reflected by the minimal values of the spread-out function. For all our experimental data sets, the minimum of the long-range spread-out function is lower than the short-range one (see Figure 6). Although some points exhibit a relative match between the two minima, i.e both models can be invoked to describe the resulting correlations, they are associated to relatively high effective temperatures \proofs{(first 4 image indexes)}. Since the two models map one another in this regime, this feature was expected. However, for lower effective temperatures \proofs{(last 4-5 image indexes)}, the difference is more pronounced, and the short-range model has more \proofs{difficulties} in describing the experimental values. Therefore, when describing the final magnetic configuration of an artificial array of nano-magnets subjected to a field-demagnetization protocol, long-range dipolar interactions have to be taken into account.}

\new{Another interesting feature emerges from the shape of the $K(T/J_{\alpha \beta})$ function plot. The long range model always exhibits a distinctive minimum of $K(T/J_{\alpha \beta})$, yielding a single effective temperature that best fits the experimental data. However, due to the flat-band behavior of the short-range spin-spin correlations in the spin-ice regime, the spread-out function presents a minimum plateau, as can be seen in Figure 5. In this case, the experimental points can be slid freely along the temperature axis, without reducing the square deviation, and therefore yielding a wide range of temperatures that all fit the experimental data.}

\new{The signature of long-range dipolar interactions is even better highlighted by the nearest-neighbor charge correlations. \proofs{The experimental values taken from all 10 images} are reported in Figure 3. For points corresponding to relatively high-temperatures, i.e  $T/J_{\alpha \beta} > 1$,  there is a good mapping between the short range model correlations and the long range correlations corresponding to both the ksi and kI model. If experimental points fall in this temperature window, there is no clear difference in terms of pairwise charge correlations between the ksi and kI models, thus giving rise to an apparent universality that transcends the geometry of the nanomagnets [24]. However, as mentioned in the previous paragraph, the dipolar model offers an overall better fit, and the short-range model has severe difficulties in explaining some of our data points.}

\begin{figure}
\includegraphics[width=8cm]{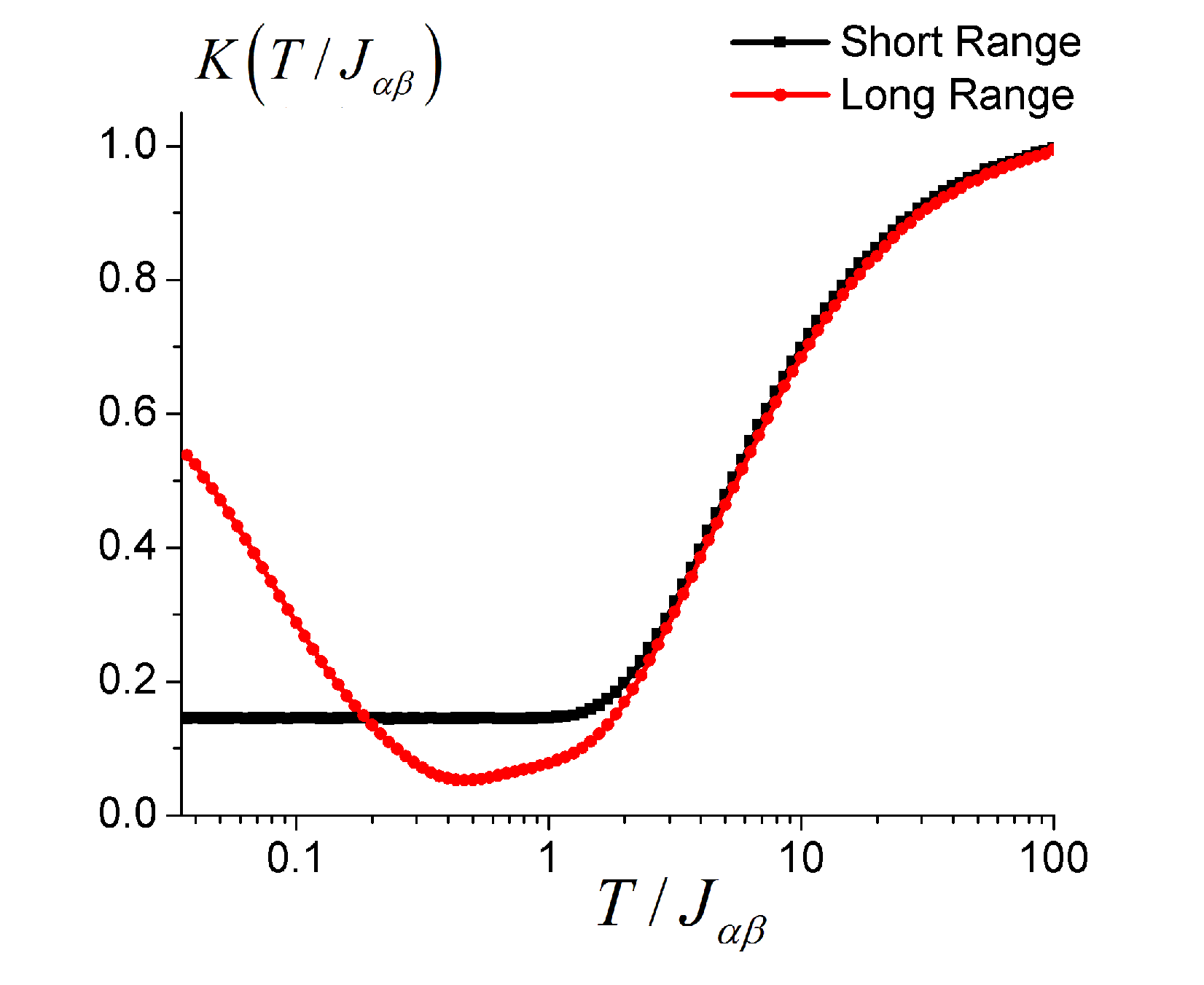}
\caption{\label{fig5}
\new{(Color online) Spread-out function computed for all Monte Carlo temperatures for both the short range (black) and the long-range (red) kI models. The experimental correlations defining this plot correspond to the tryout with the lowest effective temperature (tryout 10 - see Figure 6).}}
\end{figure}

Five of our experimental charge correlation values fall outside the theoretical standard deviations of the short-range model, making them statistically unlikely events, whereas the long range model accounts for all these values. An extreme case is presented in the inset of Figure 3, where the histogram of the $\langle Q_i.Q_{i+1} \rangle$ values expected  for $T/J_{\alpha \beta} = 0.46$ in the short-range picture is characterized by a mean value $\langle Q_{sr} \rangle = -0.116$ and a standard deviation $\sigma = 0.029$. With our experimental value of -0.021, larger than $\langle Q_{sr} \rangle + 3\sigma$�, the probability to fall into this magnetic configuration after demagnetizing the array is about 1/1000 if only nearest-neighbor interactions are considered. However, this is not the case for the dipolar model, where the experimental value can be well placed on the  $\langle Q_i.Q_{i+1} \rangle$ curve without making it a statistical extreme event. Similar features have been reported for artificial realizations of the ksi model \cite{Rougemaille2011}. Therefore, long-range dipolar interactions cannot be neglected when describing the magnetic configurations observed after demagnetizing the arrays using a field protocol.

Similar to what is observed experimentally for the multiaxial, ferromagnetic kagome spin ice, driving artificial realizations of the uniaxial, antiferromagnetic kagome Ising model into a low-temperature regime using a demagnetization protocol is challenging \cite{Rougemaille2011}. The system often remains close to the onset of the spin ice phase (i.e. $T/J_{\alpha \beta} \sim 1$). In this temperature window, a careful analysis of the pairwise spin and charge correlations over a large number of nanomagnets is required to determine \proofs{whether the short or long range model} best describes the measurements. Doing so, we find that artificial realizations of the kI model are dipolar and interactions beyond nearest-neighbors cannot be neglected. Since artificial arrays of nanomagnets are dipolar by essence, this result was expected at (very) low temperatures, as the long range part of the magnetostatic interaction differs considerably in the ksi and kI models. \new{This difference could be further emphasized experimentally by the use of thermally-active artificial spin ice structures that have been recently introduced \cite{Farhan2013a, Farhan2013b, Kapaklis2012, Zhang2013, Montaigne2014}. Since the dipolar interactions lift the degeneracy of the spin-ice manifold, such artificial arrays could be brought out of this cooperative disordered regime and further develop long-range correlations that could lead to exotic magnetic phases.} However, the importance of our work is to show that, even in the high temperature regime\proofs{,} where AC demagnetizing protocols experimentally bring the system, a full dipolar treatment is required to properly describe the measured spin and charge correlators. We can therefore assess the limits of the equivalence (universality) previously established.

\begin{figure}
\includegraphics[width=8cm]{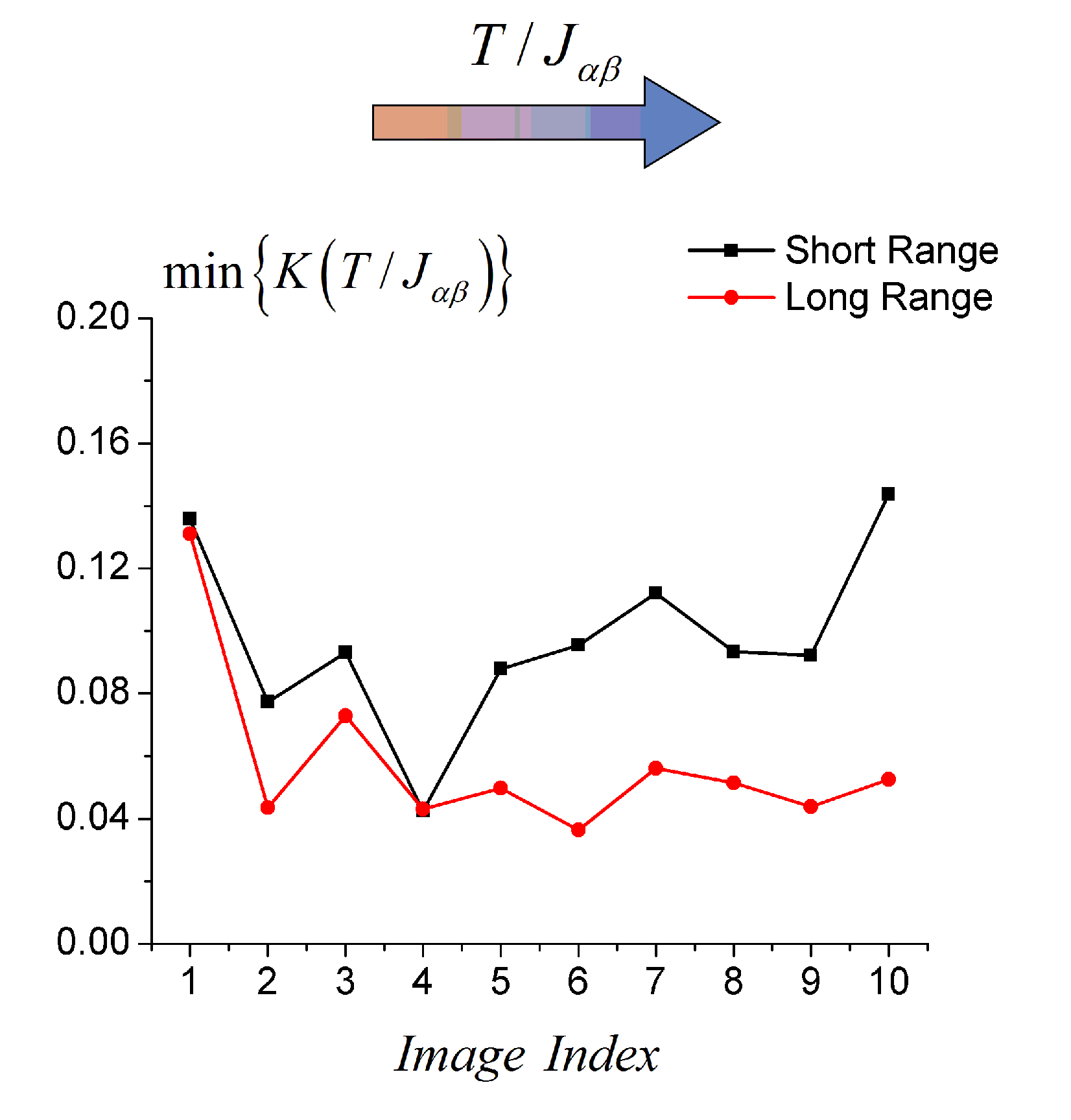}
\caption{\label{fig6}
\new{(Color online) The minimal values of the spread-out  function for all 10 datasets. In all cases, the long range model yields a better fit. The lines between the data points have no physical meaning and serve just as guides for the eye.}}
\end{figure}

This work was partially supported by the Region Lorraine and the Agence National de la Recherche through the project ANR‐12-BS04-009 `Frustrated'. I.A. Chioar acknowledges financial support from the Laboratoire d'excellence LANEF Grenoble. The authors also thank S. Le-Denmat for technical help with AFM/MFM measurements.


\begin{thebibliography}{2}
\bibitem{Note_Nico}
Within the point-dipole approximation, the coupling coefficients $J_{i j}$ between two magnetic moments $\mu \, \vec{S}_i$ and $\mu \, \vec{S}_j$ located at $\vec{r}_i$ and $\vec{r}_j$, respectively, can be expressed as: $J_{i j} = - D \, ( \, \frac{1}{|\vec{r}_{ij}|^{3}} - 3 \frac{( \vec{S}_i . \vec{r}_{ij} ) ( \vec{S}_j . \vec{r}_{ij} )}{|\vec{r}_{ij}|^{5} \, \vec{S}_i . \vec{S}_j} \, = - D \frac{1}{|\vec{r}_{ij}|^{3}},$ with $\vec{r}_{ij} = \vec{r}_i - \vec{r}_j$, and $D = ( \mu_0 / 4 \pi) \mu^2$ the strengh of the dipolar coupling.
\bibitem{Note_Ioan}
The definition of the spread-out function is a matter of choice and we have chosen to take all experimentally found correlations into account (spins and charge). However, the same final conclusion can be reached if the effective temperature is defined by taking only the charge correlations into account or the set of spin correlations.
\end{thebibliography}
\end{document}